\documentclass[longbibliography,groupedaddress,twocolumn]{revtex4-2}
\usepackage{graphicx}
\usepackage[percent]{overpic}
\usepackage{braket}
\usepackage{amsmath}
\usepackage{multirow}
\usepackage{booktabs}
\usepackage[table,xcdraw]{xcolor}
\usepackage{bm}
\usepackage{upgreek}

\usepackage{fancyhdr}
\usepackage[overlay,absolute]{textpos}
\newcommand\PlaceText[3]{
	\begin{textblock*}{10in}(#1,#2)
		#3
	\end{textblock*}
}

\begin{document}

\newcommand{\myTitle}{Gigahertz measurement-device-independent quantum key distribution using directly modulated lasers}

\newcommand{\myAuthors}{
	R.~I.~Woodward$^{1,2,*}$,
	Y.~S.~Lo$^{1, 3}$,
	M.~Pittaluga$^{1, 4}$,
	M.~Minder$^{1,5}$,
	T.~K.~Para\"{i}so$^1$,
	M.~Lucamarini$^1$,
	Z.~L.~Yuan$^1$,
	A.~J.~Shields$^1$
}

\newcommand{\myAffiliations}{
	$^1$Toshiba Europe Ltd, Cambridge, UK\\
	$^2$Quantum Communications Hub, University of York, UK\\
	$^3$Quantum Science and Technology Institute, University College London, UK\\
	$^4$School of Electronic and Electrical Engineering, University of Leeds, UK\\
	$^5$Department of Engineering, University of Cambridge, UK
}

\newcommand{\myEmail}{robert.woodward@crl.toshiba.co.uk}

\title{\myTitle}
\author{\myAuthors}
\affiliation{\myAffiliations}
\email[]{\myEmail}

\date{\today}

\begin{abstract}
Measurement-device-independent quantum key distribution (MDI-QKD) is a technique for quantum-secured communication that eliminates all detector side-channels, although is currently limited by implementation complexity and low secure key rates.
Here, we introduce a simple and compact MDI-QKD system design at gigahertz clock rates with enhanced resilience to laser fluctuations---thus enabling free-running semiconductor laser sources to be employed without spectral or phase feedback.
This is achieved using direct laser modulation, carefully exploiting gain-switching and injection-locking laser dynamics to encode phase-modulated time-bin bits.
Our design enables secure key rates that improve upon the state of the art by an order of magnitude, up to 8 bps at 54~dB channel loss and 2 kbps in the finite-size regime for 30~dB channel loss.
This greatly simplified MDI-QKD system design and proof-of-principle demonstration shows that MDI-QKD is a practical, high-performance solution for future quantum communication networks.
\end{abstract}

\maketitle

\PlaceText{12mm}{8mm}{npj Quantum Information \textbf{7}, 58 (2021); https://doi.org/10.1038/s41534-021-00394-2}

\section*{Introduction}
Quantum key distribution (QKD) is a maturing technology that enables distant communication with information-theoretic security~\cite{Bennett1984, Gisin2002}.
The development of such systems is particularly important at present as advances in quantum computation pose a growing threat to current security models based on public-key cryptography.
While QKD offers unbreakable security in theory, the practical deviation of real-world components from their ideal properties can introduce side channels that could be exploited by an eavesdropper. 
For example, a number of attacks on single-photon detectors (SPDs) have already been reported, as these are typically the most vulnerable components~\cite{Lo2014}.

To eliminate these side channels, measurement-device-independent QKD (MDI-QKD)~\cite{Lo2012} (see also~\cite{Braunstein2012}) has emerged as a promising new approach.  
Here, the two communicating users (Alice and Bob) independently encode and transmit light pulses to a central node (Charlie), which interferes them and measures the result using SPDs.
This measurement indicates the correlation between bits, but not their values, which remain known to only Alice and Bob.
This makes the protocol secure even if Charlie acts as a malicious party. 
To emphasise this important detail, Charlie is often denoted `untrusted', meaning that no trust has to be put on the intermediate node to guarantee the full security of the MDI-QKD protocol. 

Numerous recent works have demonstrated the potential of MDI-QKD, showcasing transmission distances up to 404~km~\cite{Yin2016}, field trials~\cite{Tang2015d,Tang2016b,Rubenok2013} and photonic-chip implementations~\cite{Semenenko2020}, alongside ongoing theoretical advancements~\cite{Ma2012b,Wang2013d,Chan2014a,Curty2014,Pirandola2015,Zhou2016b}.
However, the majority of MDI-QKD demonstrations to date have operated with low (sub-100~MHz) clock rates, which has limited achievable bit rates~\cite{Xu2019}.
While gigahertz-clocked MDI-QKD systems have very recently been realised using polarisation encoding~\cite{Comandar2016, Wei2019}, for practical deployments, time-bin encoding is preferred as it provides inherent immunity to depolarisation and polarisation-mode dispersion.
An additional obstacle to date has been the implementation complexity of MDI-QKD systems---in particular, the requirement for additional servo links between Alice and Bob as it is believed that instabilities in independent laser sources make high-quality two-photon interference impossible without stabilisation by active feedback~\cite{Rubenok2013}.
Significant progress is therefore still needed to develop practical high-bit-rate MDI-QKD systems which are sufficiently compact and robust for real-world environments.

\begin{figure*}[tbph]
	\includegraphics{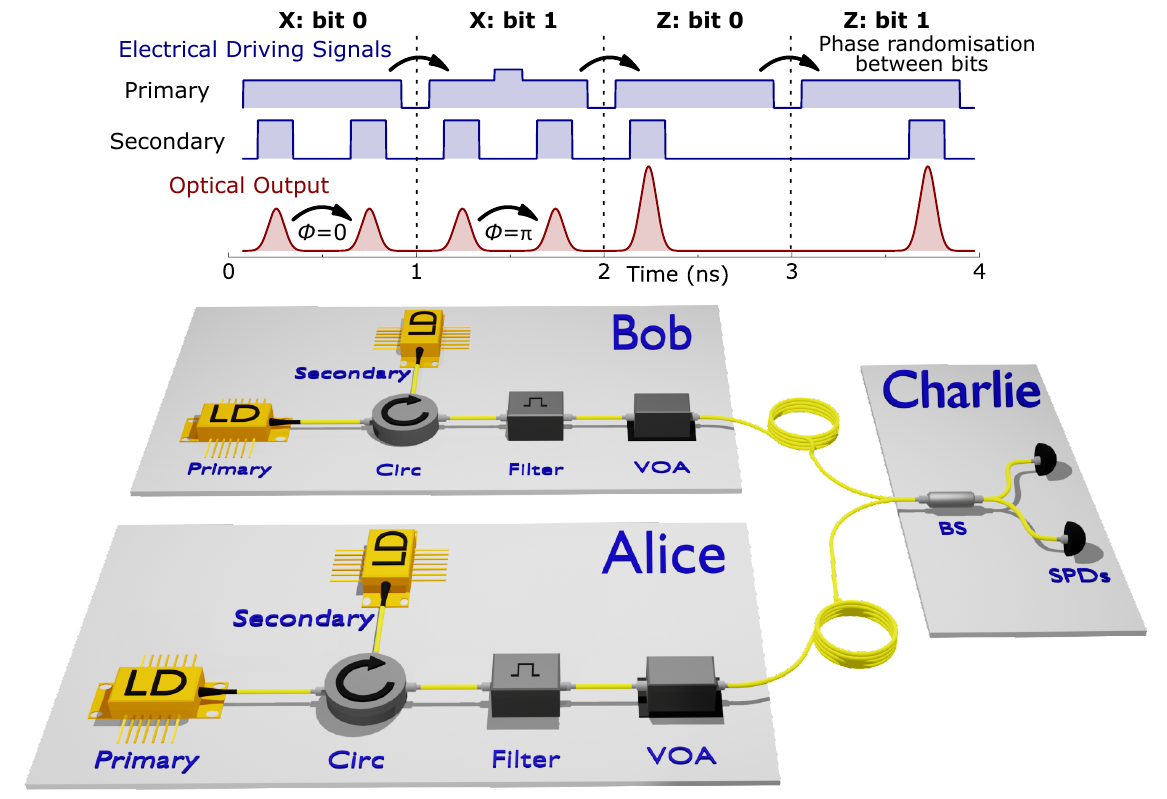}
	\caption{MDI-QKD system experimental setup. Acronyms: laser diode (LD), circulator (Circ), variable optical attenuator (VOA), beamsplitter (BS), single-photon detector (SPD). The inset illustration (top) shows the electrical signals applied to the lasers and the corresponding optical output for each basis/bit encoding.}
	\label{fig:setup}
\end{figure*}

In this Letter, we demonstrate a decisive advance in this direction with a compact, simplified, gigahertz-clocked MDI-QKD system that improves upon state-of-the-art secure key rates by up to an order of magnitude.
Much of the experimental complexity of previous MDI-QKD systems is eliminated by using directly modulated lasers to implement bit encoding, which allows us to remove the phase / polarisation modulators that have previously been required.
We also disprove the requirement for active feedback by showing that increased clock rates significantly improve the phase error robustness against free-running laser fluctuations, paving the way to practical deployment.

\section*{Results}

\subsection*{Bit Encoding}
In single-photon time-bin-encoded MDI-QKD, Alice and Bob prepare qubits in one of two bases: $Z$-basis states with the bit value encoded by the position of a photon in either the early ($\ket{0}$) or late ($\ket{1}$) time bin of the clock period; or $X$-basis states comprising a coherent superposition of a photon across both time bins with bit value encoded in the phase between them, $(\ket{0} + e^{i\phi}\ket{1})/\sqrt{2}$, where $\phi$ is 0 or $\pi$ for the $\ket{+}$ and $\ket{-}$ states, respectively.
Here, we implement the decoy-state MDI-QKD protocol, using attenuated pulsed laser sources to generate weak coherent states, with decoy states to bound single-photon events~\cite{Lo2012,Wang2013d,Zhou2016b}.
The generated pulses from each user must be indistinguishable in all degrees of freedom, such that high-visibility two-photon Hong-Ou-Mandel (HOM) interference between phase-randomised time bins can occur at Charlie~\cite{Rarity2005,Comandar2016b}.
This places exacting requirements on the light sources and previously reported MDI-QKD experiments have required multiple intensity and phase (or polarisation) modulators to achieve this~\cite{Yin2016,Rubenok2013,Tang2015d,Tang2016b,Semenenko2020}.

In our MDI-QKD system (illustrated in Fig.~\ref{fig:setup}), we generate encoded bit states using directly modulated injection-locked gain-switched lasers.
The optical injection locking technique enables precise control of phase between pulses, in addition to increasing laser modulation bandwidth and decreasing pulse chirp and jitter~\cite{Lau2009, Yuan2014a}.
Both Alice and Bob employ identical transmitter designs, including a pair of distributed feedback (DFB) lasers in a primary/secondary (i.e. master/slave) arrangement with thermoelectric controllers for temperature stabilisation.
The `primary' laser of each pair is gain switched at 1~GHz clock rate, periodically bringing the laser above threshold so that each primary pulse acquires a random optical phase.
These pulses are then injected through a circulator into the `secondary' laser, which is gain switched at 2~GHz to generate two secondary pulses within each primary pulse, effectively forming the early and late time bins for each clock period.
As secondary pulses are seeded by the injected primary pulses, they inherit the phase of the injected light and the output pulse jitter and chirp are significantly reduced~\cite{Yuan2016, Roberts2018c}.
The secondary laser output is then filtered to remove any spurious spontaneous emission and attenuated to the single-photon level (see Methods for further details).

For $Z$-basis encoding, the electrical signal to the secondary laser is patterned, selectively turning the laser on in only one time bin of each bit period. 
For $X$-basis encoding, an amplitude perturbation is selectively added in the middle of the electrical driving signal applied to the primary laser, changing the laser dynamics and slightly adjusting the output phase (Fig.~\ref{fig:setup} inset). 
As two secondary pulses are seeded by each primary pulse, this electrically controlled phase perturbation offers a versatile solution for adjusting the relative phase between the secondary pulses, which represent early and late time bins within a coherent state.
In both bases, the global phase of each coherent state is random as seeded by a new primary pulse each time, thus satisfying the security condition of MDI-QKD based on decoy states~\cite{Lo2012}.
The duration of optical pulses in our setup is measured to be 75~ps and their phase randomisation is confirmed by interfering pulses from adjacent coherent states using an unbalanced Mach-Zehnder interferometer and measuring the output intensity probability distribution (see Supplementary Note 1).

\begin{figure}[tb]
	\centering
	\sffamily
	\includegraphics{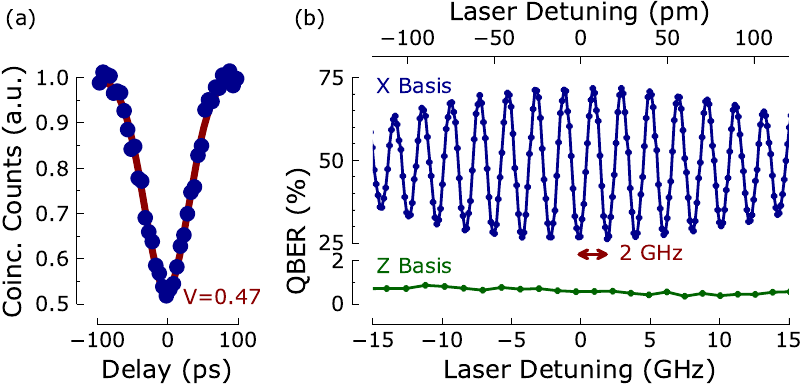}
	\rmfamily
	\caption{Interference between pulses from Alice \& Bob. (a) Hong-Ou-Mandel interference as a function of temporal delay. (b) Measured QBER as a function of spectral detuning between Alice \& Bob's lasers.}
	\label{fig:hom}
\end{figure}

The setup in Charlie comprises a 50\% beam splitter followed by two superconducting nanowire SPDs with time-tagging electronics to measure the result of interfering pulses from Alice and Bob.
By careful temporal alignment of the laser-driving waveforms applied in the transmitters (generating pulses with mean photon number commensurate with our later QKD experiments), we achieve a HOM interference visibility, $V$, of 47\% (Fig.~\ref{fig:hom}a) at 1~GHz clock rate, close to the 50\% theoretical maximum.

To perform MDI-QKD, Alice and Bob randomly choose a basis and bit value, then send encoded weak coherent states to Charlie.
By measuring the time-resolved output of the interference, these states are projected in the so-called Bell basis where certain time-resolved coincident detections indicate a successful Bell state measurement (BSM).

\subsection*{Overcoming Laser Frequency Drift}
While the optical frequency difference between Alice and Bob's lasers must be small to ensure spectral indistinguishability, there is an additional subtlety that places even tighter requirements on minimising this difference: $X$-basis phase misalignment.
Since $X$-basis bits are encoded in the phase between early and late time bins, a laser optical frequency detuning $\Delta f$ (not to be confused with the pulse generation rate) between Alice and Bob will misalign the phase basis, increasing the quantum bit error rate (QBER). 
Specifically, QBER varies sinusoidally with phase error $\Delta\phi$, where $\Delta\phi$ is related to time-bin spacing $\Delta t$ by:
\begin{equation}
\label{eqn:phase_error1}
\Delta\phi = 2 \pi \Delta t \Delta f.
\end{equation} 

The impact of this effect is observed experimentally by measuring the $X$-basis QBER as the wavelengths of Bob's lasers are varied through thermal tuning (Fig.~\ref{fig:hom}b), while the wavelengths of Alice's lasers are unchanged.
Fringes are observed as the phase error varies periodically between 0 and $\pi$, with 2~GHz period (as expected since our time-bin spacing is 500~ps). 
As our transmitters generate weak coherent states, the contribution from multiphoton events sets the theoretical minimum QBER as 25\% in the $X$-basis~\cite{Lo2012}.
Fig.~\ref{fig:hom}b also shows that the fringe amplitude decreases with increasing detuning, due to reducing spectral indistinguishability causing reduced HOM interference visibility.
The $Z$-basis QBER is also measured and observed to show no dependence upon spectral detuning (Fig.~\ref{fig:hom}b), as expected since interference is not required to produce $Z$-basis  Bell-state correlations.

\begin{figure}
	\includegraphics{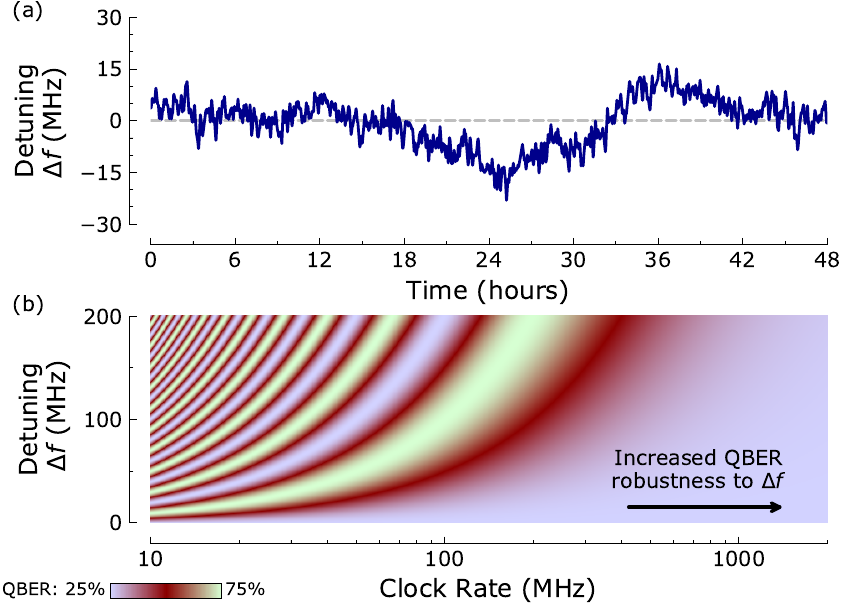}
	\caption{Laser detuning analysis. (a) Experimental time-varying spectral detuning $\Delta f$ between Alice's and Bob's free-running DFB lasers. (b) Computed $X$-basis QBER arising from phase error as a function of detuning and clock rate (assuming equally spaced time bins), showing that a higher clock rate permits greater spectral detuning without significant QBER increase.}
	\label{fig:beat}
\end{figure}

To understand the practical importance of this effect on MDI-QKD systems, we consider the typical spectral detuning that can occur between independent free-running lasers (e.g.\ arising from uncontrolled thermal/mechanical perturbations) with only thermoelectric temperature stabilisation.
The outputs from Alice's and Bob's lasers, which are typical telecommunication-grade devices, are interfered on a photodetector and the beat note is recorded over a 48-hour period.
As shown in Fig.~\ref{fig:beat}(a), the spectral difference varies by more than 30~MHz.

The impact of this laser frequency detuning is first considered for an example MDI-QKD clock rate of 75~MHz (as used by numerous prior MDI-QKD demonstrations~\cite{Tang2015d,Yin2016,Tang2016b}), which corresponds to time-bin separation, $\Delta t = 6.67$~ns assuming equally spaced time bins within each bit period.
Using Eqn.~\ref{eqn:phase_error1}, we compute that this intrinsic spectral fluctuation causes over 0.4$\pi$ phase error, which could increase the $X$-basis QBER by as much as 17\%.
This is a significant increase which could prevent the generation of a secure key.
Indeed, this problem has necessitated the addition of servo links and active feedback to stabilise laser wavelengths in previous studies, for example using either additional lasers as a phase reference~\cite{Tang2015d} or a regular beat note measurement to control an additional included frequency shifter~\cite{Valivarthi2015}.
It has even been stated that long-term high-visibility HOM interference is impossible to achieve without active stabilisation~\cite{Rubenok2013}.
An alternative recent solution was the use of a more complex protocol, known as reference-frame-independent (RFI) MDI-QKD, preparing states in 3 conjugate bases where only one basis is well defined, although with the added complexity of requiring the preparation of 3 basis states rather than 2~\cite{Wang2015d}.

We solve this problem in our system very simply, however, by using a gigahertz clock rate so that the time-bin separation is significantly reduced to 500~ps.
Using Eqn.~\ref{eqn:phase_error1}, the 30~MHz laser drift is calculated to induce 0.03~$\pi$ phase error here, which corresponds to only 0.11\% $X$-basis QBER increase.
This demonstrates the effect of spectral fluctuations on our system is negligible, eliminating the need for active feedback.
To further illustrate this phenomenon, Fig.~\ref{fig:beat}(b) shows the theoretically computed QBER increase due to this phase error with respect to clock rate and spectral difference.
Fringes (shown by red bands, where the worst-case scenario 50\% QBER indicates random selection) are observed as a function of spectral difference, which increase in period with increasing clock rate.
Therefore, greater clock rates permit a large spectral difference between Alice's and Bob's lasers before the QBER increases significantly.
This analysis reveals that smaller temporal separation between time bins makes the system intrinsically more robust against laser wavelength fluctuations.

\begin{figure}[tb]
	\centering
	\sffamily
	\includegraphics{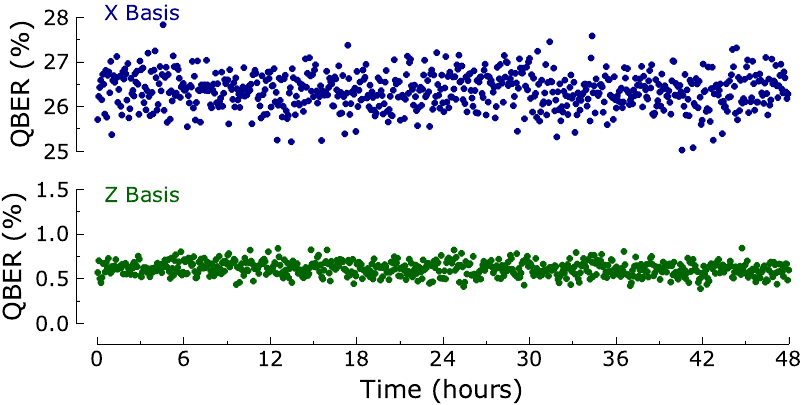}
	\rmfamily
	\caption{Measured QBER in $X$-basis and $Z$-basis showing long-term stability.}
	\label{fig:qber_wrt_time}
\end{figure}

To assess the long-term stability of our setup, the QBER is recorded over a 48-hour period (Fig.~\ref{fig:qber_wrt_time}).
Excellent stability is observed, with 0.08\% QBER standard deviation in the $Z$-basis QBER and 0.39\% in the $X$-basis, and no long-term performance drift.
This error could be caused by residual indistinguishability of the states due to minor temporal modulation imperfections.

\subsection*{Decoy-State MDI-QKD Protocol}
We use a decoy-state MDI-QKD protocol to bound single-photon yields and errors~\cite{Wang2013d,Zhou2016b}.
Specifically, a four-intensity decoy state protocol~\cite{Zhou2016b,Comandar2016} is chosen, comprising a single signal state ($s$) in the $Z$-basis and three decoy states ($u$, $v$, and $w$) in the $X$-basis.
Pulses in the $Z$ basis are used to distil a secure key while $X$-basis events are used to bound information leakage to a potential eavesdropper.

As well as presenting results in the asymptotic regime, we also apply two different methods to account for finite-size effects, both with final security parameter, $\epsilon<10^{-10}$.
The first assumes that statistical fluctuations follow a Gaussian distribution~\cite{Ma2012b}---a widely used assumption e.g. Refs.~\cite{Liu2013d,Tang2014d,Comandar2016}---while the second makes no such assumptions and applies the Chernoff bound to guarantee composable security against even the most general eavesdropper attacks~\cite{Curty2014,Comandar2016} (see Supplementary Note 2 for further details and key rate analysis).

Finally, we note that all protocol parameters (i.e. state intensities and preparation probabilities) have been carefully optimised.
This was achieved by simulating the protocol including measured component parameters \cite{Chan2014c,Comandar2016} and numerically optimising for maximum secure key rate (see Supplementary Note 3).

\subsection*{MDI-QKD System Performance}
We demonstrate proof-of-principle gigahertz-clocked MDI-QKD at various communication channel lengths, emulated using a variable optical attenuator and assuming the channel comprises ultra-low-loss fibre (0.16~dB~km$^{-1}$).
In each case, the total channel loss is comprised of equal losses for the Alice-to-Charlie link and the Bob-to-Charlie link.
Optimised parameters and tabulated gains / QBER measurements are presented in the Supplementary Tables.
QBERs as low as 0.55\% are recorded in the $Z$-bases and as low as 26.6\% in the $X$-bases, close to the 25\% theoretical minimum.

\begin{figure*}
	\includegraphics{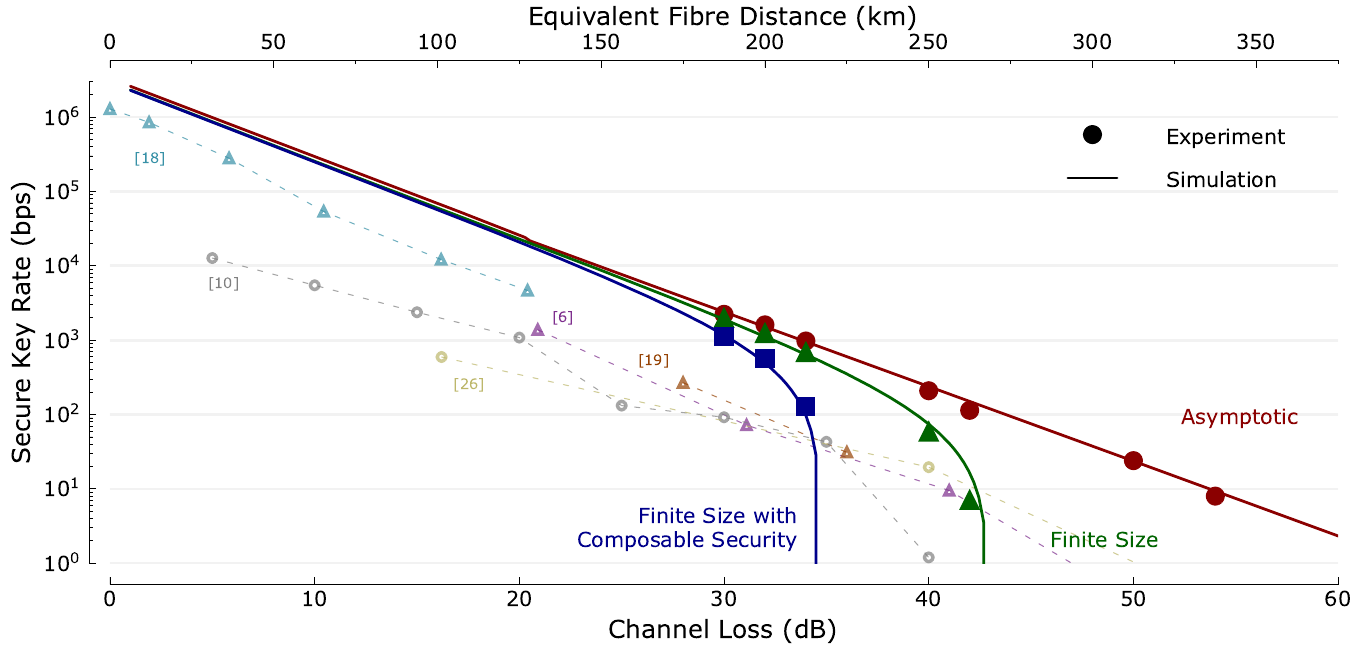}
	\caption{MDI-QKD system performance. Secure key rate with respect to total channel loss (and equivalent distance in ultra-low loss 0.16~dB~km$^{-1}$ fibre) is shown for the asymptotic (circular markers), finite size (triangular markers) and finite size with composable security (square markers) analyses. Our experimental data (filled markers) are in good agreement with numerical simulations (solid lines) based on our experimental parameters. State-of-the-art results from literature are also shown using unfilled markers~\cite{Comandar2016, Yin2016, Wei2019, Semenenko2020, Valivarthi2015}. 
	}
	\label{fig:skr}
\end{figure*}

Measurements for all possible states are then processed to estimate achievable secure key rates, as shown in Fig.~\ref{fig:skr} together with other state-of-the-art proof-of-principle MDI-QKD demonstrations~\cite{Valivarthi2015,Yin2016,Comandar2016,Wei2019,Semenenko2020}.
When accounting for finite-size effects, a secure key can be generated up to 42~dB total loss (263~km). 
Secure key rates of 1.97~kbps and 58 bps are recorded at 30 dB (188~km) and 40~dB (250~km), respectively.
Key rates are reduced slightly when applying the more restrictive finite-size security analysis, which ensures composable security against the most general attacks, but still achieving 1.12~kbps at 30~dB (188~km) and operation up to 36~dB.
Figure~\ref{fig:skr} shows that even higher key rates and larger channel losses are possible in the asymptotic regime, with 8 bps key rate measured for 54~dB channel loss.
However, for practical MDI-QKD implementations, it is essential to consider finite-size effects.
Compared to prior state-of-the-art results~\cite{Xu2019}, our GHz MDI-QKD system design enables significantly improved key rates in the finite-size regime by approximately an order of magnitude for channel losses up to 42~dB.

Performance at channel losses below 30~dB could not be measured experimentally due to saturation of our SPDs at high count rates, although we include simulation results which are well fitted by the experimental data.
This suggests that with higher maximum count rate SPDs (e.g. avalanche photodiodes), our system could achieve $>$1~Mbps MDI-QKD at metropolitan / access network distances (e.g. $<$25~km).

\section*{Discussion}
In this work we have demonstrated a simplified approach to MDI-QKD by employing directly modulated semiconductor lasers at gigahertz clock rates, enabling high-speed quantum communication without detector vulnerabilities.
Coherent states are encoded by varying the electrical waveforms that are applied to laser diodes, exploiting gain switching and optical injection locking techniques to carefully manipulate  intensity and phase.
While previous works have employed injection-locking and gain-switching of lasers to generate pulses for MDI-QKD applications~\cite{Comandar2016, Wei2019}, a fixed amplitude driving signal has always been used meaning subsequent phase/polarisation modulators were required to actually encode bit information onto the pulses.
Our approach uses a similar injection-locked gain-switched laser arrangement, but applies a carefully modulated electrical waveform to harness the laser dynamics such that time-bin phase-encoded states are produced directly, eliminating the need for subsequent bit modulation.
In addition to removing significant complexity, this approach is highly versatile as the clock rate can be varied in-situ.
We demonstrate that with a 1~GHz clock rate and fully optimising all system parameters with a finite-size security analysis, record secure key rates are achieved.
We have also shown how the high clock rate relaxes the influence of inherent laser spectral variations, further reducing complexity by remaining robust despite the use of independent, free-running DFB lasers.

Compared to other recent high-clock-rate MDI-QKD implementations that employed polarisation encoding, our time-bin encoding scheme offers advantages due to its greater immunity to polarisation-related decoherence: the probability of a photon moving from one time bin to another is very small, whereas rotation between polarisation states using polarisation encoding is unavoidable in long-distance fibres.
In addition, with our directly modulated transmitter design, it is simple to achieve low ($<1\%$) QBER for the key-generating basis since bits are encoded by on-off keying the laser source, thus ensuring maximum orthogonality between bit 0 and 1 states.
Our simplified transmitter design and elimination of wavelength-stabilising servo links therefore mitigates the barriers of high system size, cost and complexity, which has thus far limited practical MDI-QKD deployments.

The injection-locked gain-switched laser design is also well-suited for photonic chip-scale implementations~\cite{Paraiso2019,Semenenko2020}, offering a route to ultra-compact mass-manufacturable QKD systems without detector vulnerabilities.
Our results suggest that $>$1~Mbps secure key rates are possible over metropolitan/access network distances of up to 25~km and we note that the MDI-QKD architecture is well suited for building QKD networks in a star topology~\cite{Tang2016b}, with the high-cost SPDs in a central location and many users possessing only a compact low-cost transmitter. 
Recent results have also demonstrated multiplexing of classical and quantum signals in MDI-QKD, even revealing that MDI-QKD offers improved resilience to Raman noise compared to conventional point-to-point QKD~\cite{Valivarthi2019}.
MDI-QKD is therefore posed to play a valuable role in the real-world exploitation of quantum communications.

It is also interesting to consider extending the directly modulated laser scheme to higher clock rates, which could lead to increased key rates and even greater permissible spectral detuning between Alice and Bob without a phase error penalty.
There are two challenges to this goal: firstly, higher speed electronics are needed to generate electrical  patterns with shorter modulation features; and secondly, laser sources are required with greater modulation bandwidth.
The latter aspect is the primary obstacle, with current commercial laser diodes limited to $\sim$10~GHz bandwidth---therefore, increasing clock rates beyond a few GHz using current laser sources will lead to inter-pulse correlations~\cite{Yoshino2018}, which violates the requirement for phase randomised bits.
Future advances in laser technology offer to circumvent this limit, however, with recent impressive progress in higher-modulation bandwidth lasers (e.g. $>$10s~GHz) through careful device engineering~\cite{Yamaoka2021} and further optimisation of optical feedback and injection locking~\cite{Liu2020}.

It should be noted that our progress is compatible with the new embodiment of the MDI-QKD concept that uses single (rather than co-incident) event detection, known as twin-field QKD (TF-QKD)~\cite{Lucamarini2018, Minder2019}.
This enables quantum-secured communication over remarkable distances, exceeding 500~km of fibre~\cite{Chen2019a}, provided there is phase coherence between light arriving from Alice and Bob.
This requirement could be fulfilled by estimating locally in Alice's and Bob's modules the phase difference between the primary lasers and a common reference light distributed by means of frequency dissemination techniques.

In conclusion, we have demonstrated a system design for quantum-secured communications using injection-locked gain-switched laser sources, which is immune from detector side-channel attacks. 
Our approach enabled time-bin encoding at gigahertz clock speeds in a simple, compact setup, improving upon state-of-the-art MDI-QKD key rates by around an order of magnitude, with up to 8 bps at 54~dB channel losses in the asymptotic regime and up to 7 bps at 42~dB  accounting for finite-size effects.
With high-performance yet a simple design, MDI-QKD has strong potential for practical communication applications.

\section*{Methods}

\subsection*{Transmitters}
Each transmitter comprises two telecommunications-grade DFB lasers with 1550.12~nm nominal centre wavelength and an integrated thermoelectric cooler.
The primary laser includes an internal isolator to prevent back reflection, while the secondary laser is unisolated, enabling optical injection through a circulator.
Each laser is driven by an electrical waveform from a signal generator, combined with a DC bias using a bias-tee.
The primary (secondary) laser is gain switched at 1~GHz with 85\% duty cycle (2~GHz with 30\% duty cycle) and the primary/secondary waveforms are carefully temporally aligned to ensure two secondary pulses are generated within each injected primary pulse.
The RF amplitude, DC bias and (thermally tuned) wavelength of the primary and secondary lasers are carefully adjusted to optimise the quality of the generated pulses: ensuring $>$99\% phase coherence transfer by injection from primary to secondary, phase randomisation between each coherent state and indistinguishability between the two transmitters such that high-visibility HOM can be observed.
We achieve optimum performance with 45~mA bias current applied to the primary lasers, 22~mA bias to the secondary lasers and $\sim$150~$\upmu$W injection power, where the free-running wavelength of the primary and secondary lasers are within 15~GHz to ensure good injection locking when the secondary laser is seeded.

To encode pulses in the $Z$-basis, the secondary laser electrical signal is patterned to only lase in the early or late time bin of each primary pulse.
To encode pulses in the $X$-basis, the secondary electrical signal is a continuous 2~GHz waveform and additional voltage modulation is added to the primary waveform: for a bit 0, no modulation is applied to the primary waveform, so both pulses in the bit have the same phase; for a bit 1, an amplitude modulation feature with 100~ps width is applied to the primary electrical signal during the time between the secondary pulses being generated---the required amplitude to achieve a $\pi$-phase shift was determined empirically.
The pulsed secondary laser output is then passed through a bandpass spectral filter (12~GHz bandwidth) to eliminate any spurious emission and an attenuator to reduce the mean photon flux to the required level.
All lasers are independently temperature stabilised to 0.01~deg~C using a PID controller and once parameters have been set, the transmitters operate stably without need for manual adjustment or feedback servo links between the users and Charlie.

\subsection*{Detection}
The detectors at Charlie are superconducting nanowire SPDs with around 73\% detection efficiency and 40~Hz dark counts.
Time taggers are used to record detection events with 100~ps resolution and a 300~ps gate is applied to each 500~ps time bin measurement to suppress noise.
Due to the polarisation-dependence of SPDs and need for polarisation indistinguishability for good HOM interference, a polarisation controller and polarising beam splitter (PBS) are also included at Charlie.
Polarisation is manually adjusted to minimise the output from the PBS rejection port, then the polarisation remains stable without further adjustment throughout our experiment. 
To compensate polarisation drifts in practical deployments, the PBS rejection port output could simply be used as a feedback signal to automatically tune the polarisation controllers, as widely employed for real-world QKD systems and without needing a servo channel link.
There is an additional 1.4~dB loss introduced in Charlie by the fibre-optic components - this is accounted for in simulations and is not lumped into the total channel loss.

\subsection*{MDI-QKD Measurements}
Using linear optics, only two of the four possible Bell states can be measured, here comprising the `triplet' state $\ket{\Psi^+}$, corresponding to detection events in both the early and late time bin of a given detector, and the `singlet' state $\ket{\Psi^-}$, corresponding to detection events in the early time bin of detector 1 and the late time bin of detector 2 (or the late time bin of detector 1 and the early time bin of detector 2).
Due to SPD deadtime ($>20$~ns), $\ket{\Psi^+}$ cannot be measured in practice; therefore, all BSMs in this work correspond to $\ket{\Psi^-}$.

Each state encoding is measured separately in order to assess their gain and QBER, using the VOA to set the intensity for decoy states (in a practical real-time deployment, an intensity modulator could be added before the communications channel to adjust the photon flux at GHz speeds for decoy states).
Prior to performing QKD with independent randomised encodings at Alice and Bob, a pre-determined alignment pattern is sent with the resulting measurements announced by Charlie, such that Alice and Bob can adjust their timing delays to ensure good synchronisation.
A reference clock is electrically distributed between the time tagger and waveform generators, so once aligned the system does not require adjustment.
In practically deployed implementations, the clock could be distributed optically though multiplexing with classical communication service channels.

\section*{Acknowledgments}
We acknowledge EPSRC funding through the UK Quantum Communications Hub (Grant EP/M013472/1) and from the European Union Horizon 2020 Research and Innovation Programme (Grant 857156 ``OPENQKD'').

\section*{Author Contributions}
RIW and YSL performed the experiments, with support from MM, MP and TKP.
RIW and YSL performed the numerical simulations, supported by ML.
ML, ZLY and AJS supervised the project.
RIW wrote the manuscript.
All authors discussed the results and commented on the manuscript.

%%%%%%%%%%%%%%%%%%%%% Supplemental Material %%%%%%%%%%%%%%%%%%%%%

\clearpage
\onecolumngrid
\section*{Supplemental Material for \myTitle}

\begin{center}
	\myAuthors
	
	\smallskip
	
	\myAffiliations
	
	\myEmail
	
	\bigskip
\end{center}

\maketitle

\section*{Supplementary Note 1: Phase Randomisation}

To satisfy the security proofs of decoy-state MDI-QKD it is important that phase is randomised between weak coherent states. 
Our setup intrinsically achieves this by the nature of gain-switching the primary laser: by periodically driving the laser below threshold each clock cycle for a sufficient time for the laser cavity to empty of photons, each pulse grows from spontaneous emission---i.e. is effectively seeded by random vacuum fluctuations.

This is confirmed by passing the unattenuated pulse train (where each pulse has 75~ps duration, as shown in Supplementary Fig.~\ref{fig:phase_rand}a) from each transmitter through an asymmetric Mach-Zehnder interferometer (AMZI) with one arm delayed to interfere consecutive coherent states.
The output intensity is measured on a photodiode and oscilloscope, then processed to form a histogram of the output intensities at the centre of 10$^5$ pulses.
The histogram (Supplementary Fig.~\ref{fig:phase_rand}b) exhibits the $1+\cos(\phi)$ shape expected for interference of pulses with uniformly distributed random relative phase $\phi$, including accounting for experimental uncertainties~\cite{Yuan2014}. 

\begin{figure}[hptb]
	\centering
	\sffamily
	\begin{overpic}{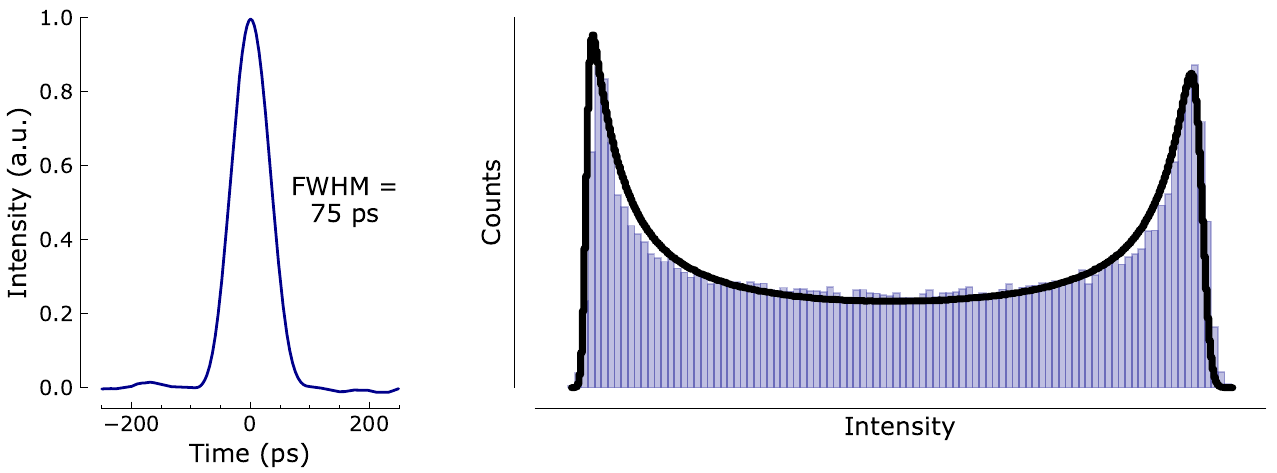}
		\put(-5, 35){ {\small (a)} }
		\put(31, 35){ {\small (b)} }
	\end{overpic}
	\rmfamily
	\caption{Characterisation of transmitter output: (a) pulse shape measured on oscilloscope; (b) histogram of pulse amplitudes after AMZI (blue bars) and simulated AMZI output accounting for experimental imperfections (black line) showing that pulse phase is randomised.}
	\label{fig:phase_rand}
\end{figure}

\section*{Supplementary Note 2: Four-Intensity Decoy State Protocol \& Key Rate Calculation}
\label{sec:theory}
We employ a 4-intensity decoy-state protocol~\cite{Zhou2016b} for MDI-QKD, described as follows.
Alice and Bob prepare weak coherent states with fluxes $\mu_i$ and $\mu_j$, respectively, encoded with a random bit value $\lbrace 0, 1 \rbrace$.
Pulses prepared in the $Z$-basis are signal states with flux $\mu=s$, which are used to generate secure key material, whereas states in the $X$-basis can be $\mu\in\lbrace u,v,w \rbrace$, corresponding to 3 possible decoy states that are used for parameter estimation to bound the single-photon yield and error rate.
States are prepared randomly with selection probability $P \in \lbrace P_Z^s, P_X^{u}, P_X^{v}, P_X^{w} \rbrace$, where the probability of encoding in the $Z$-basis is $P_Z=P_Z^s$ and in the $X$-basis is $P_X = P_X^{u} + P_X^{v} + P_X^{w}$, such that $P_Z + P_X = 1$.

To generate a secure key, Alice and Bob send their prepared states along the quantum communication channel to Charlie, who performs a Bell state measurement and announces which states resulted in a successful projection onto the singlet Bell state.
Alice and Bob then engage in classical communication to share basis information and perform sifting, parameter estimation, error correction and privacy amplification.
In practical communication, the finite sample size of states measured prior to processing means that all measurements are subject to statistical fluctuations, which thus need to be incorporated in the security analysis.

A proof-of-principle measurement is performed to highlight the secure key rate that is achievable from our gigahertz-clocked MDI-QKD system. 
The gain and QBER are measured for all possible combinations of weak coherent states, denoted $Q_\Theta^{\mu_i,\mu_j}$ and $E_\Theta^{\mu_i,\mu_j}$, respectively, for basis $\Theta \in \lbrace X, Z \rbrace$ and flux $\mu \in \lbrace s, u, v, w \rbrace$.
These measurements are then used to estimate lower-bounded single-photon yield $\underline{y}_\Theta^{1, 1}$ and upper-bounded error rate $\overline{e}_\Theta^{1,1}$, from which a lower bound of the secure key rate is computed as:
\begin{equation}
	\label{eqn:skr}
	R = \underline{q}^{1,1}_Z \left[1 - h\left(\overline{e}^{1,1}_X\right)\right] - f_{EC}~Q_{Z}^{s,s} ~h\left(E^{s,s}_Z\right) - \Delta.
\end{equation}
where $\underline{q}_Z^{1,1} = \underline{y}^{1,1}_Z \left(P_Z~s~e^{-s}\right)^2$, $h(\cdot)$ is the binary entropy function, $f_\mathrm{EC}=1.16$ accounts for error-correction inefficiency and $\Delta$ is a reduction due to finite size effects (discussed later).

We now summarise the relevant equations and our numerical approach for computing the single-photon yield and error bounds, in order to evaluate Eqn.~\ref{eqn:skr}. 
Full proofs can be found in Refs.~\cite{Ma2012b,Curty2014,Zhou2016b} and a detailed derivation of the security analysis in Ref.~\cite{Comandar2016}.

\subsection{Asymptotic Analysis}
We begin with an asymptotic analysis that neglects finite-size effects, enabling our results to be compared to other works where statistical fluctuations are not considered.
The first step is to find the single-photon yield and error rate in the $X$-basis, $\underline{y}_X^{1,1}$ and $\overline{e}_X^{1,1}$.
While analytical equations have been derived to estimate these quantities~\cite{Xu2013f}, a more accurate value that takes all experimental constraints into account can be found by following a numerical approach. 
The gain of each weak coherent state combination is related to a summation over all possible photon number states that are emitted by Alice and Bob, accounting for Poisson photon statistics.
By truncating this infinite summation, inequalities are formed which act as constraints for the numerical optimisation of $\underline{y}_X^{1,1}$, in addition to the logical requirement that $0\leq y_X^{1,1}\leq1$.
Therefore, $\underline{y}_X^{1,1}$ can be estimated by the linear program:
\begin{alignat}{2}
	& \text{minimise}   & \quad & y^{1,1}_X           \\
	~
	& \text{subject to} &       & 0 \leq y_X^{m,n} \leq 1,  \nonumber\\
	~
	& & & e^{-(\mu_i + \mu_j)} \sum_{m,n=0}^{S_\mathrm{cut}} \frac{\mu_i^m \mu_j^n}{m! n!} y^{m,n}_X  
	\geq
	\max \left\lbrace Q_X^{\mu_i,\mu_j} - \gamma_{ij}, 0  \right\rbrace,   \nonumber \\
	~
	& & & e^{-(\mu_i + \mu_j)} \sum_{m,n=0}^{S_\mathrm{cut}} \frac{\mu_i^m \mu_j^n}{m! n!} y^{m,n}_X   
	\leq 
	Q_X^{\mu_i,\mu_j},          \nonumber
\end{alignat}
where 
\begin{equation}
	\gamma_{ij} = e^{-(\mu_i + \mu_j)} \left[
	\left(
	\sum_{m=0}^{S_\mathrm{cut}} \frac{\mu_i^m}{m!}
	\right)
	\left(
	e^{\mu_j}  -   \sum_{n=0}^{S_\mathrm{cut}} \frac{\mu_j^n}{n!}	
	\right)
	+
	\left(
	e^{\mu_i}  -   \sum_{m=0}^{S_\mathrm{cut}} \frac{\mu_i^m}{m!}	
	\right)
	\left(
	\sum_{n=0}^{S_\mathrm{cut}} \frac{\mu_j^n}{n!}
	\right)
	+
	\left(
	e^{\mu_i}  -   \sum_{m=0}^{S_\mathrm{cut}} \frac{\mu_i^m}{m!}		
	\right)
	\left(
	e^{\mu_j}  -   \sum_{m=0}^{S_\mathrm{cut}} \frac{\mu_j^n}{n!}		
	\right)
	\right]
\end{equation}
and the program is solved for all $n,m \leq S_\mathrm{cut}$ ($S_\mathrm{cut}=15$ is an arbitrary integer that defines the maximum summation term) and $\mu_i,\mu_j \in \lbrace u, v, w \rbrace$.
In practice, we solve the linear programming problems using the revised simplex method.

Following similar arguments, an additional linear program can be derived to estimate $\overline{e}_X^{1,1}$:
\begin{alignat}{2}
	& \text{maximise}   & \quad & b^{1,1}_X            \\
	~
	& \text{subject to} &       & 0 \leq b_X^{m,n} \leq 1,  \nonumber\\
	~
	& & & e^{-(\mu_i + \mu_j)} \sum_{m,n=0}^{S_\mathrm{cut}}
	\frac{\mu_i^m \mu_j^n}{m! n!} b^{m,n}_X
	\geq 
	\max \left\lbrace B_X^{\mu_i,\mu_j}  - \gamma_{ij}, 0  \right\rbrace,     \nonumber \\
	~	
	& & & e^{-(\mu_i + \mu_j)} \sum_{m,n=0}^{S_\mathrm{cut}} \frac{\mu_i^m \mu_j^n}{m! n!} y^{m,n}_X e_X^{m,n}  \leq B_X^{\mu_i,\mu_j},           \nonumber
\end{alignat}
where bit error rate (BER) $B_X^{\mu_i,\mu_j} = Q_X^{\mu_i,\mu_j} E_X^{\mu_i,\mu_j}$ and $b_X^{m,n}=y_X^{m,n}e_X^{m,n}$.
The single-photon error rate is extracted from the numerical solution by: $\overline{e}_X^{1,1} = \overline{b}_X^{1,1} / \underline{y}_X^{1,1}$.
Finally, we note that in the asymptotic limit, $\underline{y}_Z^{1,1}=\underline{y}_X^{1,1}$ and $\Delta=0$, enabling these numerically optimised terms to be inserted into Eqn.~\ref{eqn:skr} to estimate the secure key rate.

\subsection{Finite-Size Analysis}
We now advance our analysis to account for statistical fluctuations in the data sample, starting with a widely used assumption that the fluctuations follow a Gaussian distribution~\cite{Ma2012b}. 
A fluctuation function is defined: $F(\zeta,n) = n / \sqrt{\zeta}$ where $n$ is the number of standard deviations which are summed over to quantify the statistical error in each measured value~\cite{Ma2012b} (here, we use $n=7$ to limit the failure probability to $<10^{-10}$~\cite{Comandar2016}).
Using this function, the linear programming problems to find $\underline{y}_X^{1,1}$ and $\overline{e}_X^{1,1}$ in the finite-size regime can be written as:

\begin{alignat}{2}
	& \text{minimise}   & \quad & y^{1,1}_X           \\
	~
	& \text{subject to} &       & 0 \leq y_X^{m,n} \leq 1,  \nonumber\\
	~
	& & & e^{-(\mu_i + \mu_j)} \sum_{m,n=0}^{S_\mathrm{cut}} \frac{\mu_i^m \mu_j^n}{m! n!} y^{m,n}_X   
	\geq
	\max \left\lbrace
	Q_X^{\mu_i,\mu_j} \left[1 - F(N_X^{\mu_i,\mu_j}Q_X^{\mu_i,\mu_j}, 7) \right] - \gamma_{ij}, 0
	\right\rbrace,         \nonumber \\
	~
	& & & e^{-(\mu_i + \mu_j)} \sum_{m,n=0}^{S_\mathrm{cut}} \frac{\mu_i^m \mu_j^n}{m! n!} y^{m,n}_X   
	\leq
	\min \left\lbrace 
	Q_X^{\mu_i,\mu_j}\left[1 + F(N_X^{\mu_i,\mu_j}Q_X^{\mu_i,\mu_j}, 7) \right], 1	
	\right\rbrace,
	\nonumber
\end{alignat}
and
\begin{alignat}{2}
	& \text{maximise}   & \quad & b^{1,1}_X            \\
	~
	& \text{subject to} &       & 0 \leq b_X^{m,n}  \leq 1,  \nonumber\\
	~
	& & & e^{-(\mu_i + \mu_j)} \sum_{m,n=0}^{S_\mathrm{cut}} \frac{\mu_i^m \mu_j^n}{m! n!}
	b^{m,n}_X
	\geq
	\max \left\lbrace
	Q_X^{\mu_i,\mu_j} \left[1 - F(N_X^{\mu_i,\mu_j}B_X^{\mu_i,\mu_j}, 7) \right] - \gamma_{ij}, 0
	\right\rbrace,         \nonumber \\
	~
	& & & e^{-(\mu_i + \mu_j)} \sum_{m,n=0}^{S_\mathrm{cut}} \frac{\mu_i^m \mu_j^n}{m! n!} b^{m,n}_X 
	\leq
	\min \left\lbrace 
	Q_X^{\mu_i,\mu_j}\left[1 + F(N_X^{\mu_i,\mu_j} B_X^{\mu_i,\mu_j}, 7) \right], 1	
	\right\rbrace,
	\nonumber
\end{alignat}
where $N_X^{\mu_i,\mu_j}$ is the number of states sent in the finite sample in the $X$-basis with fluxes $\mu_i$ and $\mu_j$.
Finally, the finite-size $Z$-basis single-photon yield is found as: $\underline{y}_Z^{1,1} = \underline{y}_X^{1,1} - \theta$  where $\theta$ takes into account possible fluctuations converting from $\underline{y}_X^{1,1}$ to $\underline{y}_Z^{1,1}$ and is fixed as $1.5\times10^{-6}$~\cite{Comandar2016}. These single-photon quantities can then be inserted into Eqn.~\ref{eqn:skr} to obtain the secure key rate.

\subsection{Finite-Size Analysis with Composable Security}
Finally, we discuss a more advanced finite-size analysis that relaxes assumption about the distribution of noise fluctuations and guarantees composable security against even coherent eavesdropper attacks~\cite{Curty2014,Comandar2016}.
This is based on applying the multiplicative Chernoff bound to quantities in the measurement sample.

We start by bounding the number of states transmitted by the users $N_\Theta^{\mu_i,\mu_j}$, for each basis ($\Theta$) and photon flux ($\mu_i, \mu_j$) combination.
These quantities are disclosed during classical communication after the whole sample is measured, but due to the users' independent choices of basis and intensity, they are fluctuating quantities that must be bounded in the finite-size regime.
By defining the function $g(x,y)=\sqrt{x \ln(y^{-2})}$, we can express the bounds such that  $\overline{N}_\Theta^{\mu_i,\mu_j}$ ($\underline{N}_\Theta^{\mu_i,\mu_j}$) is guaranteed to be greater (smaller) than $N_\Theta^{\mu_i,\mu_j}$ with probability $1 - \epsilon_0$:
\begin{align}
	\label{eqn:chern1}
	\overline{N}_\Theta^{\mu_i\mu_j}  & = N_\Theta^{\mu_i\mu_j} + g(N_\Theta^{\mu_i\mu_j}, \epsilon_0^2) \\
	\label{eqn:chern2}
	\underline{N}_\Theta^{\mu_i\mu_j} & = N_\Theta^{\mu_i\mu_j} - g(N_\Theta^{\mu_i\mu_j}, \epsilon_0^4).
\end{align}
where $\epsilon_0=4\times10^{-13}$ is used to limit the total failure probability to $<10^{-10}$.

We can then follow a similar procedure to bound the number of successful Bell state measurement counts, $C_\Theta^{\mu_i\mu_j}$, and number of erroneous Bell state counts, $EC_X^{\mu_i\mu_j}$.
(The quantities $C_\Theta^{\mu_i\mu_j}$ and $EC_\Theta^{\mu_i\mu_j}$ were directly measured in our experiment, however, they have been processed for displaying in Tables~II \& III as the gain and QBER by normalising them to to the number of transmitted states: $Q_\Theta^{\mu_i\mu_j}=C_\Theta^{\mu_i\mu_j} / N_\Theta^{\mu_i\mu_j}$ and $E_\Theta^{\mu_i\mu_j}=EC_\Theta^{\mu_i,\mu_j} / C_\Theta^{\mu_i,\mu_j}$).
By replacing $N$ in Eqns.~\ref{eqn:chern1}--\ref{eqn:chern2} with $C$ and $EC$, we find the bounds $\overline{C}_\Theta^{\mu_i\mu_j}$ \& $\underline{C}_\Theta^{\mu_i\mu_j}$ and $\overline{EC}_\Theta^{\mu_i\mu_j}$ \& $\underline{EC}_\Theta^{\mu_i\mu_j}$, respectively.

These bounded quantities can then be used to compute bounded gains and bit error rates:
\begin{align}
	\overline{Q}_\Theta^{\mu_i\mu_j}  & = \overline{C}_\Theta^{\mu_i\mu_j} /  \underline{N}_\Theta^{\mu_i\mu_j} \\
	~
	\underline{Q}_\Theta^{\mu_i\mu_j}  & = \underline{C}_\Theta^{\mu_i\mu_j} /  \overline{N}_\Theta^{\mu_i\mu_j} \\
	~
	\overline{B}_\Theta^{\mu_i\mu_j}  & = \overline{EC}_\Theta^{\mu_i\mu_j} /  \underline{N}_\Theta^{\mu_i\mu_j} \\
	~
	\underline{B}_\Theta^{\mu_i\mu_j}  & = \underline{EC}_\Theta^{\mu_i\mu_j} /  \overline{N}_\Theta^{\mu_i\mu_j}
\end{align}
which are used to reformulate the linear programming problem:
\begin{alignat}{2}
	& \text{minimise}   & \quad & y^{1,1}_X           \\
	~
	& \text{subject to} &       & 0 \leq y_X^{m,n} \leq 1,  \nonumber\\
	~
	& & & e^{-(\mu_i + \mu_j)} \sum_{m,n=0}^{S_\mathrm{cut}} \frac{\mu_i^m \mu_j^n}{m! n!} y^{m,n}_X  
	\geq
	\max \left\lbrace \underline{Q}_X^{\mu_i,\mu_j} - \gamma_{ij}, 0  \right\rbrace,   \nonumber \\
	~
	& & & e^{-(\mu_i + \mu_j)} \sum_{m,n=0}^{S_\mathrm{cut}} \frac{\mu_i^m \mu_j^n}{m! n!} y^{m,n}_X   
	\leq 
	\min \left\lbrace \overline{Q}_X^{\mu_i,\mu_j} , 1 \right\rbrace,         \nonumber
\end{alignat}
and
\begin{alignat}{2}
	& \text{maximise}   & \quad & b^{1,1}_X           \\
	~
	& \text{subject to} &       & 0 \leq b_X^{m,n}  \leq 1,  \nonumber\\
	~
	& & & e^{-(\mu_i + \mu_j)} \sum_{m,n=0}^{S_\mathrm{cut}} \frac{\mu_i^m \mu_j^n}{m! n!} b^{m,n}_X  
	\geq
	\max \left\lbrace \underline{B}_X^{\mu_i,\mu_j} - \gamma_{ij}, 0  \right\rbrace,   \nonumber \\
	~
	& & & e^{-(\mu_i + \mu_j)} \sum_{m,n=0}^{S_\mathrm{cut}} \frac{\mu_i^m \mu_j^n}{m! n!} b^{m,n}_X 
	\leq 
	\min \left\lbrace \overline{B}_X^{\mu_i,\mu_j} , 1 \right\rbrace.         \nonumber
\end{alignat}

These problems can be solved to obtain $\underline{y}_X^{1,1}$ and $\overline{b}_X^{1,1}$, from which it follows $\overline{e}_X^{1,1} = \overline{b}_X^{1,1} / \underline{y}_X^{1,1}$ and $\underline{y}_Z^{1,1} = \underline{y}_X^{1,1} - \theta$, as discussed previously.
Finally, to compute the key rate using Eqn.~\ref{eqn:skr}, we follow the approach in Ref.~\cite{Comandar2016}, defining the finite-size correction factor as $\Delta=300.5 / N_\mathrm{tot}$ and redefining $q_Z^{1,1}$ as:
\begin{equation}
	q_Z^{1,1} = \frac{1}{N_\mathrm{tot}} \max \left\lbrace
	|s^2  e^{-2s} \underline{y}^{1,1}_Z \underline{N}_Z - g(s^2  e^{-2s} \underline{y}^{1,1}_Z \underline{N}_Z, \epsilon_0)|, 0
	\right\rbrace,
\end{equation}
where $N_\mathrm{tot}$ is the total number of states transmitted in the sample.

\section*{Supplementary Note 3: Parameter Optimisation}
For optimum QKD system performance, it is important to carefully select all operating parameters in order to maximise the secure key rate.
While general best-practice trends are known, for example, biasing basis selection to prepare the majority of states in the $Z$ (key-generating) basis, the exact optimal parameters vary for each experimental implementation and also depend on the channel loss.
Therefore, a full optimisation procedure is necessary to obtain the best possible results. 

In the four-state decoy-state MDI-QKD protocol, we are required to optimise the intensities of $s$, $u$, $v$ and $w$, in addition to the probabilities of the users choosing to prepare and transmit each state. 
The probabilities of state preparation are the same for each user and the requirement for unity sum of probabilities removes one degree of freedom.
Additionally, to simplify the problem, we also fix the vacuum intensity to $w=2\times 10^{-4}$.
This still leaves six parameters to optimise in our system, which is non-trivial and beyond the scope of a simple exhaustive parameter space search approach.

\begin{figure}[b]
	\centering
	\sffamily
	\begin{overpic}{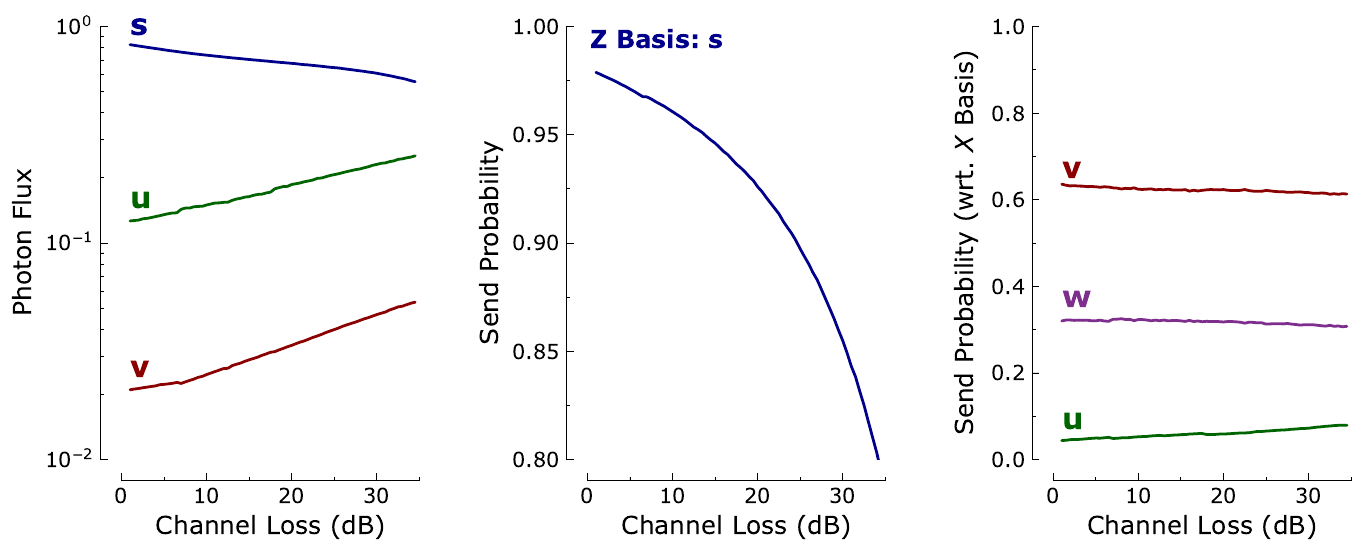}
		\put(-5, 40){ {\small (a)} }
		\put(31, 40){ {\small (b)} }
		\put(65, 40){ {\small (c)} }
	\end{overpic}
	\rmfamily
	\caption{Numerically optimised parameters for our MDI-QKD system (including finite size effects, with composable security) with respect to total channel losses: (a) state intensities ($w$ was fixed as 2$\times 10^{-4}$); (b) probability of encoding bit in Z basis (signal state $s$); (c) probability of encoding a $u$, $v$ or $w$ decoy state in the X basis (note that probabilities are plotted relative to the probably of encoding in the X basis).}
	\label{fig:optimal_params}
\end{figure}

We therefore perform a full numerical optimisation by noting that the problem is convex and using a local optimisation algorithm, bounded to experimentally practical intensities and probabilities in the range $(0,1)$.
The algorithm is provided with a fitness function to maximise, comprising the secure key rate calculation in the finite-size regime with composable security where the gain and QBER are determined by simulating the protocol using the six optimisation variables in addition to experimentally measured component values.
Supplementary Fig.~\ref{fig:optimal_params} shows the optimised parameters, which were then used in our experiments (see main text), resulting in strong agreement between theoretically modelled and measured key rates.

\clearpage

\section*{Supplementary Tables}

\begin{figure}
	\includegraphics{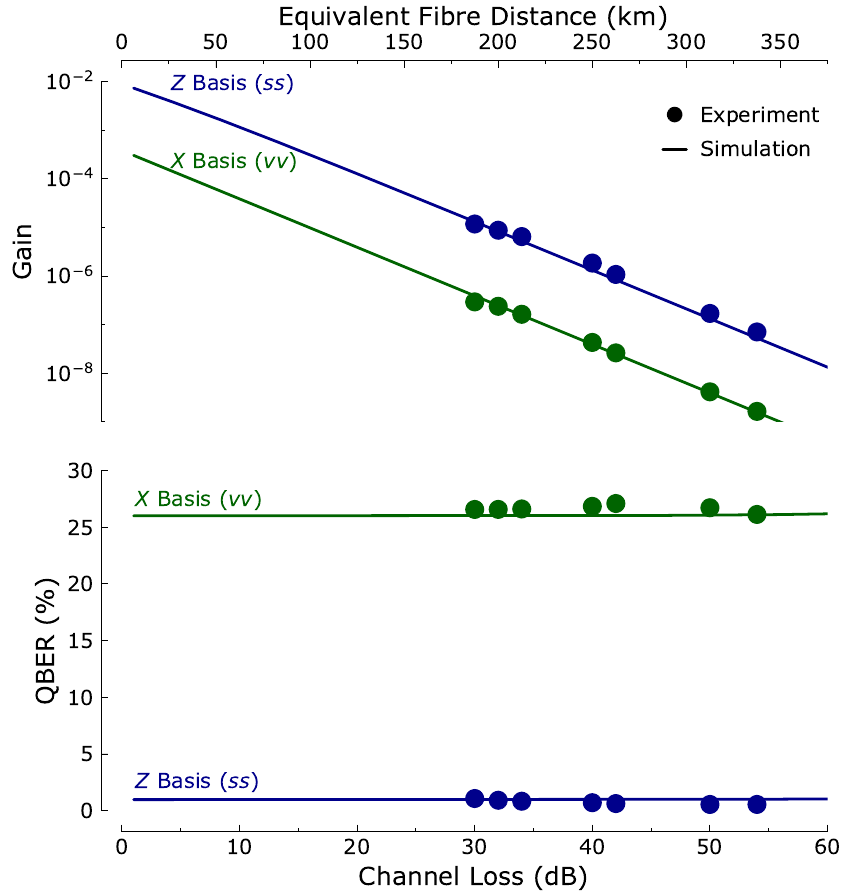}
	\caption{Gain and QBER for the X-basis (decoy state $v$) and Z-basis (signal state $s$), as a function of total channel loss (equivalent fibre distance assuming ultra-low loss 0.16~dB km$^{-1}$ fibre is also shown). Experimental data (circles) are in good agreement with numerical simulations (lines) based on our experimental parameters.}
	\label{fig:gainloss}
\end{figure}

Supplementary Tables~I--III report detailed experimental results for our proof-of-principle MDI-QKD system at various channel loss values. 
Supplementary Table~I presents the computed secure key rates using the three security analyses considered in this work, based on processing the raw experimental measurements obtained in the $Z$-basis (Table~II) and $X$-basis (Table~III).
Supplementary Fig.~\ref{fig:gainloss} plots a selection of this data to show the measured gain and QBER for the cases where both parties send a bit either in the $Z$ basis ($s$ state) or the $v$ state in the $X$ basis (discussed in main text).

%%%%%%%%%%%%%%%%%%%%%%%%%%%%%%%%%%%%
% TABLE FORMATTING
% Define centered column with fixed width
\newcolumntype{x}[1]{>{\centering\arraybackslash\hspace{0pt}}p{#1}}
% Add height to rows
\setlength\extrarowheight{4 pt}
% Add spacing to columns
\setlength{\tabcolsep}{0.2cm}

\begin{table}[hptb]	
	\begin{tabular}{@{}cccc@{}}
		\toprule
		\multirow{3}{*}{\textbf{Channel Loss}} & \multicolumn{3}{c}{\textbf{Secure Key Rate, $\bm{R}$ (bps)}}                                                          \\ 
		& \multirow{2}{*}{\textbf{Asymptotic}} & \multirow{2}{*}{\textbf{Finite Size}} & \textbf{Finite Size with}    \\
		&                                      &                                       & \textbf{Composable Security} \\ \cmidrule(r){1-4}
		\textbf{30 dB (188 km)}                       & 2228                                 & 1971                                  & 1118                         \\
		\textbf{32 dB (200 km)}                       & 1607                                 & 1227                                  & 564                          \\
		\textbf{34 dB (213 km)}                       & 975                                  & 681                                   & 130                          \\
		\textbf{40 dB (250 km)}                       & 209                                  & 58                                    & --                           \\
		\textbf{42 dB (263 km)}                       & 114                                  & 7                                     & --                           \\
		\textbf{50 dB (313 km)}                       & 24                                   & --                                    & --                           \\
		\textbf{54 dB (338 km)}                       & 8                                    & --                                    & --    \\                      
		\bottomrule
	\end{tabular}
	\label{tab:skr}
	\caption{Experimental secure key rates $R$, with respect to total channel attenuation (dB). In parentheses, we also report the equivalent distance in km, calculated by assuming ultra-low loss 0.16~dB km$^{-1}$ fibre.}
\end{table}

\begin{table}[hptb]	
	\label{tab:Z}
	\textbf{Z-Basis}\\ 
	~\\
	\begin{tabular}{@{}ccccc@{}}
		\toprule
		\multirow{2}{*}{\textbf{Channel   Loss}} & \multirow{2}{*}{\textbf{Gain, $\bm{Q_Z^{s,s}}$}} & \multirow{2}{*}{\textbf{QBER, $\bm{E_Z^{s,s}}$}} & \multirow{2}{*}{\textbf{Flux, $\bm{s}$}} & \multirow{2}{*}{\textbf{Prob., $\bm{P_Z^s}$}} \\
		&                                              &                                             &                                &                                         \\ \midrule
		\textbf{30 dB (188 km)}                           & 1.18$\times 10^{-05}$                                     & 1.07\%                                      & 0.55                           & 85.0\%                                  \\
		\textbf{32 dB (200 km)}                           & 8.75$\times 10^{-06}$                                     & 0.92\%                                      & 0.60                            & 83.0\%                                  \\
		\textbf{34 dB (213 km)}                           & 6.48$\times 10^{-06}$                                     & 0.84\%                                      & 0.63                           & 81.0\%                                  \\
		\textbf{40 dB (250 km)}                           & 1.85$\times 10^{-06}$                                     & 0.70\%                                      & 0.63                           & 81.0\%                                  \\
		\textbf{42 dB (263 km)}                           & 1.08$\times 10^{-06}$                                     & 0.63\%                                      & 0.63                           & 81.0\%                                  \\
		\textbf{50 dB (313 km)}                           & 1.70$\times 10^{-07}$                                     & 0.55\%                                      & 0.63                           & 81.0\%                                  \\
		\textbf{54 dB (338 km)}                           & 7.07$\times 10^{-08}$                                     & 0.55\%                                      & 0.63                           & 81.0\%                                  \\ \bottomrule
	\end{tabular}
	\caption{Experimental gains and errors in the $Z$-basis. These values were computed from the number of Bell state measurements, where the total prepared state sample size was $N_\mathrm{tot}=8.64\times10^{13}$ and the number of state preparations by the users is given by $N_Z^{s,s} = P_Z P_Z N_\mathrm{tot}$. The QBER is observed to decrease slightly with increasing channel attenuation, which is attributed to reduced measurement jitter in SNSPDs at lower received count rates.}
\end{table}

\begin{table}[hptb]
	\textbf{X-Basis}\\  
	~\\
	\begin{tabular}{@{}ccccccccccc@{}}
		\toprule
		\multirow{2}{*}{\textbf{Channel   Loss}} & \multicolumn{4}{c}{\textbf{Gain,   $\bm{Q_X^{\mu_i,\mu_j}}$}}                             & \multicolumn{4}{c}{\textbf{QBER,   $\bm{E_X^{\mu_i,\mu_j}}$}}                              & \multirow{2}{*}{\textbf{Flux, $\bm{\mu}$}} & \multirow{2}{*}{\textbf{Prob., $\bm{P_X^{\mu}}$}} \\ \cmidrule(lr){2-9}
		& \textbf{}           & \textit{\textbf{u}} & \textit{\textbf{v}} & \textit{\textbf{w}} & \textbf{}           & \textit{\textbf{u}} & \textit{\textbf{v}} & \textit{\textbf{w}} &                                &                                             \\ \cmidrule(r){1-1} \cmidrule(l){10-11} 
		\multirow{3}{*}{\textbf{30 dB (188 km)}} & \textit{\textbf{u}} & 6.41$\times 10^{-06}$            & 2.44$\times 10^{-06}$            & 1.70$\times 10^{-06}$            & \textit{\textbf{u}} & 26.9\%              & 36.5\%              & 48.5\%              & 0.24                           & 1.0\%                                       \\
		& \textit{\textbf{v}} & 2.51$\times 10^{-06}$            & 2.94$\times 10^{-07}$            & 7.77$\times 10^{-08}$            & \textit{\textbf{v}} & 36.5\%              & 26.6\%              & 48.2\%              & 0.047                          & 9.3\%                                       \\
		& \textit{\textbf{w}} & 1.79$\times 10^{-06}$            & 7.59$\times 10^{-08}$            & 7.32$\times 10^{-12}$            & \textit{\textbf{w}} & 48.5\%              & 47.9\%              & 32.5\%              & 0.0002                         & 4.7\%                                       \\
		\multicolumn{11}{c}{\textbf{}}                                                                                                                                                                                                                                                                          \\
		\multirow{3}{*}{\textbf{32 dB (200 km)}} & \textit{\textbf{u}} & 4.32$\times 10^{-06}$            & 1.67$\times 10^{-06}$            & 1.14$\times 10^{-06}$            & \textit{\textbf{u}} & 26.6\%              & 35.4\%              & 48.5\%              & 0.25                           & 1.2\%                                       \\
		& \textit{\textbf{v}} & 1.78$\times 10^{-06}$            & 2.38$\times 10^{-07}$            & 6.42$\times 10^{-08}$            & \textit{\textbf{v}} & 35.4\%              & 26.6\%              & 48.2\%              & 0.053                          & 10.3\%                                       \\
		& \textit{\textbf{w}} & 1.19$\times 10^{-06}$            & 5.80$\times 10^{-08}$            & 4.82$\times 10^{-12}$            & \textit{\textbf{w}} & 48.5\%              & 48.0\%              & 43.0\%              & 0.0002                         & 5.5\%                                       \\
		\multicolumn{11}{c}{\textbf{}}                                                                                                                                                                                                                                                                          \\
		\multirow{3}{*}{\textbf{34 dB (213 km)}} & \textit{\textbf{u}} & 2.53$\times 10^{-06}$            & 9.95$\times 10^{-07}$            & 6.66$\times 10^{-07}$            & \textit{\textbf{u}} & 26.5\%              & 35.0\%              & 48.5\%              & 0.24                           & 1.3\%                                       \\
		& \textit{\textbf{v}} & 1.05$\times 10^{-06}$            & 1.63$\times 10^{-07}$            & 4.56$\times 10^{-08}$            & \textit{\textbf{v}} & 34.5\%              & 26.6\%              & 48.6\%              & 0.056                          & 11.4\%                                       \\
		& \textit{\textbf{w}} & 7.02$\times 10^{-07}$            & 3.87$\times 10^{-08}$            & 3.75$\times 10^{-12}$            & \textit{\textbf{w}} & 48.4\%              & 48.5\%              & 44.6\%              & 0.0002                         & 6.3\%                                       \\
		\multicolumn{11}{c}{\textbf{}}                                                                                                                                                                                                                                                                          \\
		\multirow{3}{*}{\textbf{40 dB (250 km)}} & \textit{\textbf{u}} & 7.29$\times 10^{-07}$            & 2.73$\times 10^{-07}$            & 1.78$\times 10^{-07}$            & \textit{\textbf{u}} & 26.8\%              & 35.0\%              & 48.6\%              & 0.24                           & 1.3\%                                       \\
		& \textit{\textbf{v}} & 2.97$\times 10^{-07}$            & 4.30$\times 10^{-08}$            & 1.10$\times 10^{-08}$            & \textit{\textbf{v}} & 35.2\%              & 26.8\%              & 48.1\%              & 0.056                          & 11.4\%                                       \\
		& \textit{\textbf{w}} & 1.95$\times 10^{-07}$            & 1.06$\times 10^{-08}$            & 1.07$\times 10^{-12}$            & \textit{\textbf{w}} & 48.5\%              & 48.5\%              & 22.6\%              & 0.0002                         & 6.3\%                                       \\
		\multicolumn{11}{c}{\textbf{}}                                                                                                                                                                                                                                                                          \\
		\multirow{3}{*}{\textbf{42 dB (263 km)}} & \textit{\textbf{u}} & 4.47$\times 10^{-07}$            & 1.68$\times 10^{-07}$            & 1.10$\times 10^{-07}$            & \textit{\textbf{u}} & 26.9\%              & 35.0\%              & 48.6\%              & 0.24                           & 1.3\%                                       \\
		& \textit{\textbf{v}} & 1.85$\times 10^{-07}$            & 2.62$\times 10^{-08}$            & 6.73$\times 10^{-09}$            & \textit{\textbf{v}} & 35.4\%              & 27.1\%              & 48.1\%              & 0.056                          & 11.4\%                                       \\
		& \textit{\textbf{w}} & 1.18$\times 10^{-07}$            & 6.46$\times 10^{-09}$            & 6.89$\times 10^{-13}$            & \textit{\textbf{w}} & 48.7\%              & 47.7\%              & 42.9\%              & 0.0002                         & 6.3\%                                       \\
		\multicolumn{11}{c}{\textbf{}}                                                                                                                                                                                                                                                                          \\
		\multirow{3}{*}{\textbf{50 dB (313 km)}} & \textit{\textbf{u}} & 7.13$\times 10^{-08}$            & 2.61$\times 10^{-08}$            & 1.69$\times 10^{-08}$            & \textit{\textbf{u}} & 27.1\%              & 35.2\%              & 49.2\%              & 0.24                           & 1.3\%                                       \\
		& \textit{\textbf{v}} & 2.93$\times 10^{-08}$            & 4.15$\times 10^{-09}$            & 9.27$\times 10^{-10}$            & \textit{\textbf{v}} & 34.8\%              & 26.7\%              & 48.5\%              & 0.056                          & 11.4\%                                       \\
		& \textit{\textbf{w}} & 1.91$\times 10^{-08}$            & 1.04$\times 10^{-09}$            & 3.91$\times 10^{-14}$            & \textit{\textbf{w}} & 48.8\%              & 49.3\%              & 49.5\%              & 0.0002                         & 6.3\%                                       \\
		\multicolumn{11}{c}{\textbf{}}                                                                                                                                                                                                                                                                          \\
		\multirow{3}{*}{\textbf{54 dB (338 km)}} & \textit{\textbf{u}} & 2.91$\times 10^{-08}$            & 1.06$\times 10^{-08}$            & 6.82$\times 10^{-09}$            & \textit{\textbf{u}} & 26.9\%              & 33.8\%              & 48.2\%              & 0.24                           & 1.3\%                                       \\
		& \textit{\textbf{v}} & 1.18$\times 10^{-08}$            & 1.69$\times 10^{-09}$            & 4.04$\times 10^{-10}$            & \textit{\textbf{v}} & 35.4\%              & 26.7\%              & 49.1\%              & 0.056                          & 11.4\%                                       \\
		& \textit{\textbf{w}} & 7.89$\times 10^{-09}$            & 4.33$\times 10^{-10}$            & 3.91$\times 10^{-14}$            & \textit{\textbf{w}} & 49.0\%              & 46.0\%              & 49.5\%              & 0.0002                         & 6.3\%                                       \\
		\bottomrule
	\end{tabular}
	\label{tab:X}
	\caption{Experimental gains and errors in the $X$-basis. These values were computed from the number of Bell state measurements for each state combination, where the total prepared state sample size was $N_\mathrm{tot}=8.64\times10^{13}$ and the number of state preparations by the users is given by $N_X^{\mu_i,\mu_j} = P_X^{\mu_i} P_X^{\mu_j} N_\mathrm{tot}$.}
\end{table}

\end{document}